\documentclass[aps,superscriptaddress,twocolumn,twoside,floatfix,prl,10pt]{revtex4-1}
\usepackage[english]{babel}
\usepackage{graphicx,amsmath,amssymb,amsfonts,enumerate}
\usepackage[ocgcolorlinks,colorlinks=true,linkcolor=blue,citecolor=red,linktocpage=true]{hyperref}

\newcommand{\ket}[1]{|#1\rangle}
\newcommand{\bra}[1]{\langle#1|}
\newcommand{\rg}{\mathrm{guess}}
\newcommand{\acc}{\mathrm{acc}}

\usepackage{mathptmx}

\DeclareMathOperator{\tr}{tr}

\begin{document}

\title{Robustness of Measurement, discrimination games and accessible information}
\author{Paul Skrzypczyk}
\affiliation{H. H. Wills Physics Laboratory, University of Bristol, Tyndall Avenue, Bristol, BS8 1TL, UK.}
\author{Noah Linden}
\affiliation{School of Mathematics, University of Bristol, University Walk, Bristol, BS8 1TW, UK}

\begin{abstract}
We introduce a way of quantifying how informative a quantum measurement is, starting from a resource-theoretic perspective. This quantifier, which we call the robustness of measurement, describes how much `noise' must be added to a measurement before it becomes completely uninformative. We show that this geometric quantifier has operational significance in terms of the advantage the measurement provides over guessing at random in an suitably chosen state discrimination game. We further show that it is the single-shot generalisation of the accessible information of a certain quantum-to-classical channel. Using this insight, we also show that the recently-introduced robustness of coherence is the single-shot generalisation of the accessible information of an ensemble. Finally we discuss more generally the connection between robustness-based measures, discrimination problems and single-shot information theory.
\end{abstract}

\maketitle

%\textit{Introduction---}
\section{Introduction}
Although quantum states provide a complete description of a physical system at a given time, it is through the process of measurement that classical information about the state of the system is obtained. How much information is obtained depends upon the nature of the measurement made. Intuitively, some measurements are more informative than others, depending on how much correlation can be generated between the measurement outcomes and the state of the quantum system. Measurements which are not able to generate strong correlations -- i.e. those which lead to almost uniform measurement outcomes for all quantum states -- are naturally less informative than measurements which can lead to deterministic results. 

The study of how informative a quantum measurement is it not new. There has been a series of papers studying this question, from an information-theoretic perspective \cite{groenewold1971, massar2000, winter2001, winter2004, elron2007, buscemi2008, dallarno2011, oreshkov2011, holevo2012, holevo2013, berta2014, dallarno2014, hirche2017}. The novel approach we adopt here comes from taking a `resource-theoretic' point of view. 

In recent years there has been much interest coming from quantum information in studying quantum properties and phenomena taking a resource-theoretic perspective, whereby one treats the property or phenomenon of interest as a resource, and tries to quantify it from an operational perspective. The prototypical example of such a quantum resource theory is the theory of entanglement \cite{bennett1996a,bennett1996}, but there have been many other resource theories put forward recently, including asymmetry \cite{gour2008,marvian2013}, coherence \cite{aberg2006a,baumgratz2014a}, purity \cite{horodecki2003} thermodynamics \cite{janzing2000,brandao2013,horodecki2013}, magic states \cite{veitch2014}, nonlocality \cite{vicente2014}, steering \cite{gallego2015}, contextuality \cite{horodecki2015,abramsky2017,amaral2018} and knowledge \cite{delrio2015}. For a recent review article, see \cite{chitambar2018}.

Here we are interested in returning to the question of how informative a measurement is, starting from such a resource-theoretic perspective. A number of questions arise. Which measurements are most informative? How can we compare the informativeness of one measurement to another from this perspective?

To that end, we introduce here a way of quantifying the informativeness of a measurement, by introducing what we call the robustness of a measurement, which, roughly speaking, corresponds to the amount of `noise' that has to be added to a measurement before it ceases to be informative at all. After showing that this quantity has the usual desirable properties that one would expect from any meaningful quantifier, such as faithfulness, convexity, and non-increase under processing, we go on to show that it has a natural operational interpretation from the perspective of state discrimination, where it quantifies the best advantage that the measurement can provide over randomly guessing the state. Moreover, we also show that the robustness of measurement is naturally related to a single-shot generalisation of the accessible information of a quantum-classical channel. Thus although our starting point was different to previous work, we indeed find a close connection to many ideas already explored \cite{groenewold1971, massar2000, winter2001, winter2004, elron2007, buscemi2008, dallarno2011, oreshkov2011, holevo2012, holevo2013, berta2014, dallarno2014, hirche2017}, as one might expect. 

Using this insight, we then return to a similar quantifier that was recently introduced in the context of the resource theory of coherence/asymmetry \cite{napoli2016,piani2016}. We show that the robustness of coherence is also a single-shot generalisation, now of the accessible information of an ensemble. We believe this signals a more general connection between robustness based measures of resources, single-shot information theory, and discrimination type problems, as we discuss in the conclusions.

%\textit{Robustness of Measurements---} 
\section{Robustness of Measurement} 
Let us think about (destructive) quantum measurements starting from a resource-theoretic perspective. To that end, imagine a scenario where we have access to only one specific measuring device. That is, we have access to a box, which accepts as input an arbitrary quantum state $\rho$ (of fixed dimension $d$), and performs the measurement $\mathbb{M} = \{M_a\}_a$ with $o$ outcomes on the system, where each $M_a$ is a positive-semidefinite operator, $M_a \geq 0$, (a POVM element), which collectively sum to the identity $\sum_a M_a = \openone$. The box returns the measurement outcome $a$ with probability $p(a) = \tr[M_a \rho]$. 

Even if we only have access to the single measurement $\mathbb{M}$, we still naturally have access to another type of box, which performs a `trivial' measurement. That is, we can also consider a box $\mathbb{T} = \{T_a\}_a$ which accepts quantum states, but produces random outcomes, i.e. which returns a supposed measurement outcome $a$ with probability $q(a)$, independent of the quantum state measured. Such a measurement can be thought of as having POVM elements $T_a = q(a)\openone$. 

Using resource-theoretic language, we can think of the set of all such trivial measurements as being the \emph{free} measurements, and any measurement which is not of this form as being a \emph{resourceful} measurement -- i.e. one which genuinely performs a quantum measurement. 

It is natural to look at quantitative properties of measurements from this perspective. In particular, given a particular measurement $\mathbb{M}$, one can try to quantify to what extent it is a resourceful measurement, and to understand the physical content of such statements. Intuitively, one would hope that ideal von Neumann measurements, where each POVM element is a rank-1 projector, $M_a = \Pi_a$, should be the most resourceful measurements. 

Here we will focus on a single measure, which we term the \emph{Robustness of Measurement} (RoM), which is the analogue of robustness measures which have been widely studied in the many quantum information theory contexts, e.g.~ \cite{vidal1999,almeida2007,piani2015a,napoli2016,piani2016}. As will shall see, this particular measure has many nice properties and a compelling operational interpretation.

The RoM is defined by the minimal amount of `noise' that needs to be added to the measurement such that it becomes a trivial measurement. In particular, if instead of always performing the intended measurement $\mathbb{M}$, it was the case that sometimes a different measurement $\mathbb{N} = \{N_a\}_a$ were performed, then one is interested in the minimal probability of this other measurement which would make the overall measurement trivial. Formally,
\begin{align} \label{e:RoM}
    R(\mathbb{M}) = \min_{r,\mathbb{N},\mathbf{q}}    &\quad r \nonumber \\
    \text{s.t.}             &\quad \frac{M_a + r N_a}{1+r} = q(a) \openone \quad \forall a,  \\
                            &\quad N_a \geq 0 \quad \forall a, \quad \sum_a N_a = \openone.  \nonumber
\end{align}
In the above, the minimisation is over all `noise' measurements $\mathbb{N} = \{N_a\}_a$, and all probability distributions $\mathbf{q} = \{q(a)\}_a$. In order to have a number of convenient mathematical properties, we use the convention whereby the probability of noise is given by $r/(1+r)$. 
\section{Properties}

As is often the case for robustness based measures of a resource, the robustness of measurements has a number of desirable properties: 
\begin{enumerate}[(i)]
\item It is \emph{faithful}, meaning that it vanishes if and only if the measurement is trivial, i.e. 
\begin{equation}
R(\mathbb{M}) = 0 \iff M_a = q(a)\openone \quad \forall a.
\end{equation}
\item It is \emph{convex}, meaning that one cannot have a larger robustness by classically choosing between two measurements, i.e. for $0 \leq p \leq 1$, 
\begin{equation}
\quad\quad R(p \mathbb{M}_1 + (1-p)\mathbb{M}_2) \leq p R(\mathbb{M}_1) + (1-p) R(\mathbb{M}_2).
\end{equation}
\item It is non-increasing under any allowed measurement simulation \cite{guerini2017}. That is, given access only to a single measurement $\mathbb{M}$, we can simulate any other measurement $\mathbb{M}' = \{M'_b\}_b$ (with an arbitrary number of outcomes $b$) such that 
\begin{equation}\label{e:2}
    M_b' = \sum_a p(b|a)M_a
\end{equation}
where $p(b|a)$ form a set of conditional probability distributions (such that the matrix $[D]_{ab} = p(b|a)$ is a stochastic matrix), i.e. we do the measurement $\mathbb{M}$ and then post-process the outcome. For any such simulated $\mathbb{M'}$ we have
\begin{equation}\label{e: monotone}
R(\mathbb{M}') \leq R(\mathbb{M}).
\end{equation}
\end{enumerate}

These three properties can be easily shown, and follow the same logic as in other robustness measures. For completeness, we include the proofs in the Appendix. 

It turns out that $R(\mathbb{M})$ can be evaluated explicitly as we now show. By defining $\tilde{\mathbf{q}} = \{\tilde{q}(a)\}_a$ with $\tilde{q}(a) := (1+r)q(a)$, and using the first equality in \eqref{e:RoM} to solve for $N_a$, it can be shown that \eqref{e:RoM} can be equivalently written as
\begin{align}\label{e:primal}
    R(\mathbb{M}) = \min_{\tilde{\mathbf{q}}}   &\quad \sum_a \tilde{q}(a) - 1\nonumber \\
                \text{s.t.} &\quad \tilde{q}(a)\openone \geq M_a \quad \forall a.
\end{align}
which is explicitly in the form of a semidefinite program \cite{boyd2004}. However, by inspection the optimal solution of this optimisation problem can be identified: $\tilde{q}(a)$ will be minimised when equal to the operator norm $\|M_a\|_\infty$ (since $M_a$ is positive-semidefinite), and hence we arrive at the exact expression
\begin{equation}
    R(\mathbb{M}) = \sum_a \|M_a\|_\infty - 1.
\end{equation}
In order to be a valid POVM element it is necessary to satisfy the operator inequality $\openone \geq M_a$, from which it follows that $1 \geq \|M_a\|_\infty$ and hence $R(\mathbb{M}) \leq o-1$. Consider also the pair
\begin{align}
N_a &= \frac{\tr[M_a]\openone - M_a}{d-1},& q(a) &= \frac{1}{d}\tr[M_a]
\end{align}
which for any measurements $\mathbb{M}$ forms a valid measurement $\mathbb{N} = \{N_a\}_a$ and probability distribution $\mathbf{q} = \{q(a)\}_a$. This pair directly imply that $R(\mathbb{M}) \leq d-1$. Putting both bounds together, we thus see that
\begin{equation}
R(\mathbb{M}) \leq \min(o,d) - 1.
\end{equation}
This implies in particular that in dimension $d$ the largest robustness that can be achieved is $R(\mathbb{M}) = d-1$, which can occur only for measurements with at least $d$ outcomes. 

It is interesting to identify which measurements achieve this maximum and are maximally robustness. First, for ideal projective von Neumann measurements, $M_a = \Pi_a$, we have $\|\Pi_a\|_\infty = 1$ for all $a$, and hence $R(\mathbb{M}) = d-1$.  Consider furthermore any rank-1 measurement (with an arbitrary number of outcomes $o > d$), where $M_a = \alpha_a \Pi_a$. To be a valid measurement, $\alpha_a \geq 0$ and $\sum_a \alpha_a = d$. Such measurements are also seen to be  maximally robustness $R(\mathbb{M}') = d-1$. We will return to the meaning of this later in the paper. 

%\textit{Dual Formulation---}
%\section{Dual Formulation}  
Finally, we saw previously that the RoM can be formulated as an SDP in \eqref{e:primal}. This provides us with a second representation of the RoM, in terms of the dual formulation of the SDP \cite{boyd2004}, which will prove insightful when we come to look at the operational significance of the RoM. As demonstrated explicitly in the Appendix, strong duality holds and the dual formulation of \eqref{e:primal} is given by
\begin{align}\label{e:dual}
    R(\mathbb{M}) = \max_{\{\rho_a\}_a}   &\quad \sum_a \tr[\rho_a M_a] - 1\nonumber \\
                \text{s.t.} &\quad \rho_a \geq 0, \quad \tr[\rho_a] = 1 \quad \forall a
\end{align}
where the maximisation is now over the dual variables $\{\rho_a\}_a$ which, due to the nature of the constraints, are seen naturally to correspond to quantum states. 

As with the primal formulation, the dual can be solved explicitly by inspection. In particular, $\rho_a$ should be chosen as a projector onto any state in the eigenspace of the maximal eigenvalue of $M_a$. With such a choice, then $\tr[M_a \rho_a] = \|M_a\|_\infty$ and $R(\mathbb{M}) = \sum_a \|M_a\|_\infty - 1$ as required. 

%\textit{Operational significance of RoM---} 
\section{Operational significance}
We now turn our attention to the operational significance of the RoM. Originally we introduced it as a distance based measure of a measurement. In this section we will see that the RoM can be understood as the advantage that can be achieved in a state discrimination problem over guessing at random, if one only has use of the measurement $\mathbb{M}$.

Consider a situation where one of a set of known states $\{\sigma_x\}_x$ is prepared with probability $\mathbf{p} = \{p(x)\}_x$. Such a situation is described by an ensemble $\mathcal{E} = \{\sigma_x,p_x\}_x$. The goal, as in any state discrimination problem, is to guess which of the states has been prepared in a given round. In each round a guess $g$ will be made of which state was prepared. Our figure of merit will be the average probability of guessing correctly, i.e. $p_\rg(\mathcal{E}) = \sum_x p(x) p(g=x|x) = \sum_{x,g} p(x) p(g|x)\delta_{x,g}$, where $p(g|x)$ is the conditional probability of guessing the state $\sigma_g$, given that the state $\sigma_x$ was actually prepared. We would like to consider two situations: (i) the trivial situation where one is unable to measure the quantum states prepared. (ii) The situation where one is able to perform a fixed measurement $\mathbb{M}$ in order to produce a guess. 

In case (i), the optimal strategy is to always guess the most probable state was prepared, i.e. the state $\sigma_x$ such that $p(x) = \max_y p(y)$ (which may not be unique). If we denote by $p_{\max} = \max_x p(x)$, then in this case the probability to guess correctly is precisely $p_\rg^C(\mathcal{E}) = p_{\max}$. 

In case (ii), after measuring the state prepared, by using the measurement $\mathbb{M}$, the most general strategy is to produce a guess based upon the outcome, according to some distribution $P(g|a)$, which will lead to a guessing probability of
\begin{equation}
P_\rg^Q(\mathcal{E},\mathbb{M}) = \max_{\{P(g|a)\}}\sum_{x,a,g}p(x)\tr[\sigma_x M_a] P(g|a)\delta_{g,x}
\end{equation}
We are then interested in the state discrimination problem which maximises the ratio between these two guessing probabilities -- i.e. the discrimination problem for which having access to the measurement $\mathbb{M}$ provides the biggest advantage over having to guess at random. Formally, we are interested in the advantage
\begin{equation}
\max_{\mathcal{E}} \frac{P_\rg^Q(\mathcal{E},\mathbb{M})}{P_\rg^C(\mathcal{E})}
\end{equation}
In the Appendix we show that the advantage is specified completely by the RoM, in particular that
\begin{equation}\label{e:op sig RoM}
\max_{\mathcal{E}} \frac{P_\rg^Q(\mathcal{E},\mathbb{M})}{P_\rg^C(\mathcal{E})} = 1 + R(\mathbb{M})
\end{equation}
and that the optimal discrimination problem is to choose uniformly at random from a set of $o$ states $\{\rho^*_a\}_a$ which are optimal variables for the dual SDP \eqref{e:dual}.

Considering specific examples, for an ideal von Neumann measurement, we see that we can use this to perfectly guess which out of $d$ states were prepared, whereas without the ability to perform a measurement we would have to guess (uniformly at random), and hence the advantage is $p_\rg^Q/p_\rg^C = d$. As a second example, consider a rank-1 measurement $\mathbb{M} = \{\alpha_a \Pi_a\}_a$ For the discrimination problem with the $o$ states associated to this measurement, the guessing probability is $p_\rg^Q = d/o$, while the classical probability is $p_\rg^C = 1/o$, and again the advantage is $d$, as expected. This shows why such measurements still have maximal robustness, since they still allow for a $d$ times advantage in this context.

%\textit{Single-shot Information theory---}
\section{Single-shot Information theory} We will now demonstrate a second way of interpreting the operational significance of the RoM by making a connection to single-shot information theory, by viewing a measurement alternatively as a quantum channel which produces classical outputs. 

Given a general quantum channel, i.e. a general completely-positive and trace-preserving map $\Lambda(\cdot)$ that maps quantum states to quantum states, a basic quantity of interest is the accessible information, the maximal amount of classical information that can be conveyed by the channel \cite{wilde2013}
\begin{equation}
I^\acc(\Lambda(\cdot)) = \max_{\mathcal{E},\mathbb{D}} I(X:G)
\end{equation}
where $\mathcal{E} = \{\sigma_x,p(x)\}_x$, with $\sigma_x$ the input states to the channel, which are chosen with probability $p(x)$,  $\mathbb{D} = \{D_g\}_g$ forms a POVM which is measured on the output of the channel to produce a symbol $g$ with probability $p(g|x) = \tr[D_g \Lambda(\sigma_x)]$, and $I(X:G) = H(X) - H(X|G)$ is the classical mutual information of the distribution $p(x,g) = p(x)p(g|x)$. The accessible information thus quantifies the maximal amount of classical mutual information that can be generated between the input and output of the channel, optimising over all encodings (input ensembles) and decodings (measurements)

Since it is based upon the Shannon entropy, $I^\acc$ is an asymptotic measure of a channel. Here, we would like to consider an analogous quantity in a single-shot regime, where the channel will only be used a single time. Let us therefore consider the following single-shot variant of the mutual information \cite{ciganovic2014}
\begin{equation}
I_{\min}(X:G) = H_{\min}(X) - H_{\min}(X|G)
\end{equation}
where $H_{\min}(X) = -\log \max_x p(x)$ and $H_{\min}(X|G) = -\log \sum_g \max_x p(x,g)$ are the min-entropy and conditional min-entropy, respectively, and are the entropies associated to the guessing probability \cite{renner2005,renner2004}. We then define the accessible min-information as
\begin{equation}
I_{\min}^\acc(\Lambda(\cdot)) = \max_{\mathcal{E},\mathbb{D}} I_{\min}(X:G)
\end{equation}
where, $\mathcal{E} = \{\sigma_x, p(x)\}_x$  and $\mathbb{D} = \{D_g\}_g$ are all as before. 

A special class of quantum channels are those which correspond to measurements, i.e. quantum channels which take as input a quantum state $\rho$ and produce as output the state $\sum_a \tr[M_a\rho] \ket{a}\bra{a}$, where $\{\ket{a}\}$ forms an arbitrary orthonormal basis for the Hilbert space of the output. Let us denote by $\Lambda_\mathbb{M}$ the channel associated to the measurement $\mathbb{M} = \{M_a\}_a$ in this way. 

We show in the Appendix, that given this viewpoint, we can alternatively express the previous result that the RoM is the advantage in a state discrimination problem as
\begin{equation}
I_{\min}^\acc(\Lambda_\mathbb{M}(\cdot)) = \log(1 + R(\mathbb{M}))
\end{equation}
that is, the RoM is also equivalent to the accessible min-information of the channel $\Lambda_\mathbb{M}(\cdot)$ associated to the measurement, which is the maximal amount of min-mutual-information that can be generated between the input and output of the channel in a single use. 

%\textit{Robustness of Assymetry/coherence---}
\section{Robustness of Asymmetry and coherence}
We would now like to turn our attention to a closely related robustness-based measure that was recently introduced, the Robustness of Asymmetry (RoA) \cite{piani2016}, which has as a special case the Robustness of Coherence (RoC) \cite{napoli2016}. We will show that the above operational significance of the RoM in terms of accessible min-information of a quantum-to-classical channel has a natural analogue for the RoA and RoC, where it will also be shown to be equal to the accessible min-information of an ensemble, (for a suitably chosen ensemble), which can be thought of as a classical-to-quantum channel. 

Consider a unitary representation $U_h$ of a group $H$ \footnote{We will present here the analysis for a discrete group. It is straightforward to carry out the analysis for a continuous group also.}. A state is symmetric with respect to the group if $\rho = \frac{1}{|H|}\sum_h U_h \rho U_h^\dagger$, where $|H|$ denotes the number of elements of the group. Any state which is not symmetric is asymmetric, and is considered as a resource within the resource theory of asymmetry. The Robustness of Asymmetry (RoA) is then the minimal amount of noise that needs to be added to a state before it becomes symmetric
\begin{align}
\mathcal{A}_R(\rho) = \min_{s,\tau,\sigma} &\quad s \nonumber \\
    \text{s.t.}             &\quad \frac{\rho + s \tau}{1+s} = \sigma  \\
							&\quad \tau \geq 0, \quad \tr[\tau] = 1, \quad \sigma = \tfrac{1}{|H|}\sum_h U_h \sigma U_h^\dagger   \nonumber                       
\end{align}
In the case where the group $H$ and representation $U_h$ generate complete de-phasing with respect to a certain fixed basis, then the RoA is known as the Robustness of Coherence (RoC), and symmetry/asymmetry becomes equal to incoherent/coherent in the fixed basis. 

In \cite{napoli2016,piani2016} it was shown that the RoA has an operational interpretation as the advantage that can achieved by using an asymmetric state $\rho$ over any symmetric state in the following discrimination problem: Consider that the channel $\mathcal{U}_h(\cdot) = U_h(\cdot)U_h^\dagger$ will be applied to a state with probability $q(h)$. We will denote $\mathbf{q} = \{q(h)\}_h$. The goal is to optimally guess which channel has been applied using a fixed state $\rho$, in comparison to any symmetric state. Defining $p_\rg^Q(\mathbf{q},\rho) = \max_{\{M_y\}_y,\{P(g|y)\}} \sum_{g,h,y} q(h) \tr[U_h \rho U_h^\dagger M_y]P(g|y)\delta_{g,h}$ as the success probability for the state $\rho$, where the maximisation is over all measurements $\{M_y\}_y$ and all guessing strategies $\{P(g|y)\}_{g,y}$, and $p_\rg^S(\mathbf{q}) = \max_h q(h)$ as the maximal success probability for any symmetric state (which conveys no information about which $U_h$ was applied at all), then it was shown in \cite{napoli2016,piani2016} that
\begin{equation}
\max_{\mathbf{q}} \frac{p_\rg^Q(\mathbf{q},\rho)}{p_\rg^S(\mathbf{q})} = 1 + \mathcal{A}_R(\rho)
\end{equation}
That is, that the RoA is equal to the optimal advantage in the best state discrimination game where the states sent are created by applying a unitary  $U_h$ from the group $H$ to the state $\rho$. It was shown that the optimal game is when the states are sent uniformly at random, $q(h) = 1/|H|$.  

Here we would like to show that this result can be similarly reinterpreted as about the min-accessible information of the ensemble $\mathcal{E}_\rho = \{U_h\rho U_h^\dagger, 1/|H|\}_h$ associated to this optimal game. In particular, for an ensemble $\mathcal{E} = \{\sigma_h, q(h)\}_h$, the accessible min-information can be defined (in analogy to the accessible information \cite{wilde2013}) as
\begin{equation}
I_{\min}^\acc(\mathcal{E}) = \max_{\mathbb{M}} I_{\min}(H : Y)
\end{equation}
where $\mathbb{M} = \{M_y\}_y$ is an arbitrary POVM, and $p(h,y) = p(h)\tr[\sigma_h M_y]$. Then, it can be shown that, for ensembles of the form $\mathcal{E}_\rho = \{U_h\rho U_h^\dagger, 1/|H|\}_h$, that
\begin{equation}
I_{\min}^\acc(\mathcal{E}_\rho) = \log(1+ \mathcal{A}_R(\rho))
\end{equation}
That is, the RoA quantifies the accessible min-information of the ensemble formed by application of $\mathcal{U}_h$ to $\rho$. A short proof of this statement can be found in the Appendix. 

We finish by noting that while a measurement can be viewed as a quantum-to-classical channel, an ensemble can be thought of as a classical-to-quantum channel, taking the classical random variable $h$ to the quantum state $\sigma_h$. As such, the RoM and RoA can be seen as capturing properties of two extremal types of channels, either transforming quantum information from or to classical information. 

\section{Conclusions}
Here we have addressed the question of how informative a measurement is from a resource-theoretic perspective. We introduced a quantifier of informativeness, which we termed the Robustness of Measurement. Our main findings are to show that this quantifier exactly characterises the advantage that a measurement provides (over guessing at random) in the task of state discrimination, and, when viewing a measurement as a quantum-to-classical channel, is also equal to a single-shot generalisation of the accessible information of the channel.

Our starting point was to take a resource-theoretic perspective on measurements (similar to that taken in \cite{guerini2017}), where we are only able to perform a single measurement $\mathbb{M}$, and the free-operations are to post-process the measurement. The RoM was shown in \eqref{e: monotone} to be a \emph{monotone} in this respect, that is, non-increasing under the allowed operation. A natural question is what other monotones exist for this resource theory of measurements, and to find a complete set of monotones which characterise whether or not a measurement $\mathbb{M}'$ is a post-processing of $\mathbb{M}$ or not (i.e. to establish the partial order). In the appendix we show that a complete set of monotones exist, and are given by success probability over the set of all discrimination games \footnote{We are very grateful to F. Buscemi for point this out to us.}. That is, $\mathbb{M}'$ is a post-processing of $\mathbb{M}$ if and only if
\begin{equation}
P_\rg^Q(\mathcal{E},\mathbb{M}) \geq P_\rg^Q(\mathcal{E},\mathbb{M}') \quad \text{for all }\mathcal{E}.
\end{equation} 

There are a number of interesting directions which we leave for future work. First, we focused on a particular choice of quantifier here, which we showed had desirable properties and interesting operational significance. One can nevertheless define other quantifiers starting from the resource-theory perspective taken here, for example based upon relative entropy or other distance based measures. It would be interesting to understand how the use of other quantifiers can lead to further insights into the informativeness of a measurement.   
\begin{figure}[t!]
\begin{center}
\includegraphics[width=0.85\columnwidth]{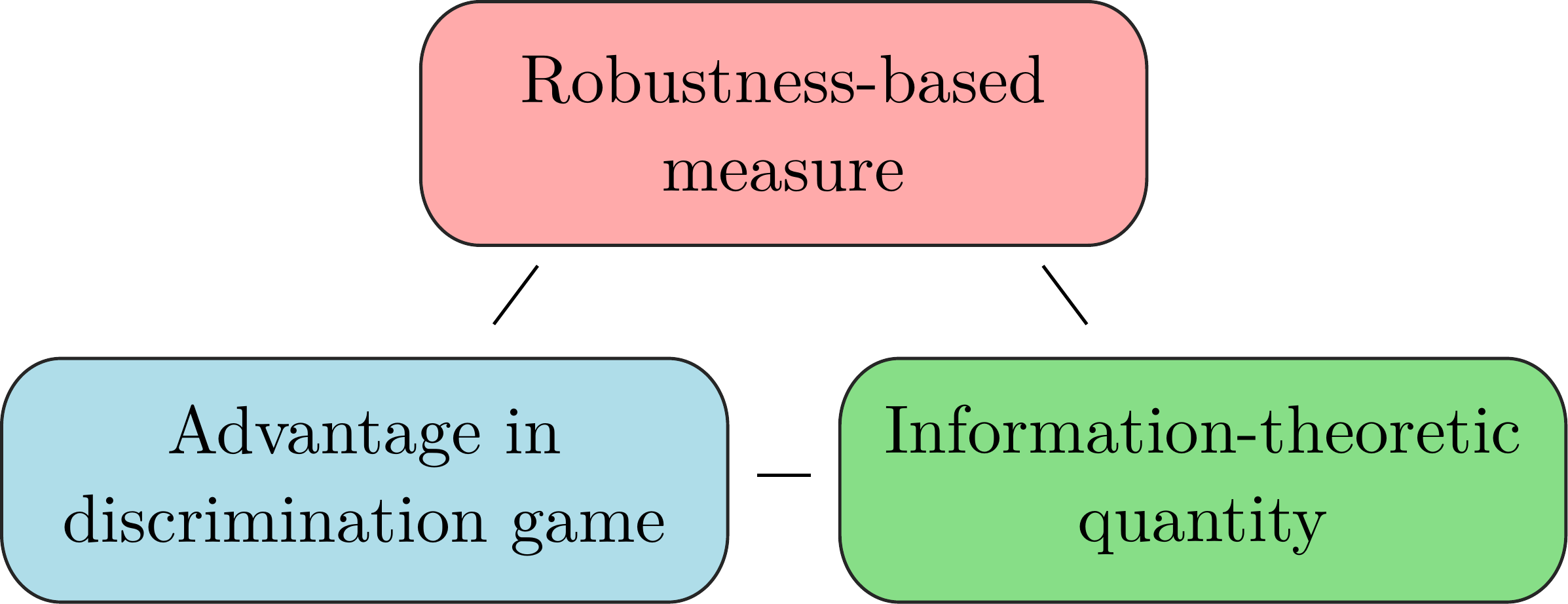}
\caption{Triangle of associations found here. A robustness based measure is found to give the optimal advantage in a suitably chosen discrimination game. This in turn is found to be equivalent to a suitably defined information theoretic quantity -- the single-shot accessible information of a suitably defined channel. It is interesting to ask how general this triangle of associations is.}
\label{f:triangle}
\end{center}
\end{figure}

Second, our work fits into a strand or research which connects robustness based measures of resources with discrimination type problems \cite{piani2015}. Although it was also known, for the case of entanglement, that the robustness-based measure was connected to single-shot information theory through the max-relative entropy of entanglement \cite{datta2009}, we believe that this is the first time where this triangle of connections has been made explicit more generally (see Fig.~\ref{f:triangle}). It would be very interesting to understand just how far this triangle of robustness-based quantifier, discrimination problem, and information-theoretic quantity  can be applied. For example, we conjecture that for any channel a single-shot accessible information is associated to a robustness of some type and moreover to a discrimination-type game. Indeed, we can ask if this triangle of associations holds very generally: given any example of a vertex of the triangle, can one find the associated other two vertices (either in the single-shot or asymptotic scenario). 

Third, here we have only considered the measurement outcomes, and not the post-measurement state. It would be interesting to extend the results here to the case of quantum instruments, where we also keep the post-measurement state. In particular, since we know there is an information-disturbance trade-off, it would be interesting to investigate this phenomenon using the robustness of measurement as the quantifier of information gain. 

Finally, in single-shot information theory it is usually necessary to introduce the possibilities of small errors -- and therefore approximations -- in order to obtain meaningful results, through the use of smoothed entropies. Here however we have not had to introduce such approximations and smoothing. It would be interesting to consider the role of approximation and smoothing when considering the measurements from this resource-theoretic perspective.  
\section{Note added}
After completing this work, the independent work of Takagi et al.~appeared online \cite{takagi2018}. In that work, the authors show a general connection between robustness-based measures for states, and discrimination games, as we conjectured here in our discussion as one link in the triangle. 
%\textit{Acknowledgements---}
\section{Acknowledgements}
PS acknowledges support from a Royal Society URF (UHQT). We thank Francesco Buscemi for insightful discussions. In particular, we thank Francesco for pointing out that a complete set of monotones can be found for measurement simulation in terms of guessing probabilities. 

\bibliography{robustness-of-meas}

\begin{appendix}
\section{APPENDICES}
\subsection{Properties of the RoM}
In this  appendix we will prove the following three properties for the RoM: (i) faithfulness: $R(\mathbb{M}) = 0 \iff M_a = q(a)\openone$ $\forall a$; (ii) convexity:  $R(p \mathbb{M}_1 + (1-p)\mathbb{M}_2) \leq p R(\mathbb{M}_1) + (1-p) R(\mathbb{M}_2)$; (iii) non-increasing under any allowed measurement simulation: $R(\mathbb{M}') \leq R(\mathbb{M})$ when $M_b' = \sum_a p(b|a)M_a$.

Property (i) follows by definition. If the measurement $\mathbb{M}$ is trivial, then $M_a = q(a)\openone$ for all $a$, and hence a solution to \eqref{e:RoM} exists with $r = 0$. If on the other hand the measurement is non-trivial, then it must be necessary to add noise in order to make it trivial, hence $R(\mathbb{M}) > 0$. 

To prove convexity, from the definition of the RoM it follows that there exists measurements $\mathbb{N}_i^*$ and probabilities $\mathbf{q}_i = \{q_i^*(a)\}_a$, for $i = 1,2$, such that 
\begin{equation}
\mathbb{M}_i = [1 + R(\mathbb{M}_i)]\mathbf{q}^*_i \openone - R(\mathbb{M}_i)\mathbb{N}_i^*
\end{equation}
where equality is understood to hold for each outcome $a$. Such decompositions are referred to as pseudo-mixtures for the measurements $\mathbb{M}_i$. It now follows immediately that 
\begin{multline}
p\mathbb{M}_1 + (1-p)\mathbb{M}_2 = p\left([1 + R(\mathbb{M}_1)]\mathbf{q}^*_1 \openone - R(\mathbb{M}_1)\mathbb{N}_1^*\right)  \\
+ (1-p)\left([1 + R(\mathbb{M}_2)]\mathbf{q}^*_2 \openone - R(\mathbb{M}_2)\mathbb{N}_2^*\right).
\end{multline}
Then, by defining
\begin{align}
\tilde{r} &= pR(\mathbb{M}_1) + (1-p)R(\mathbb{M}_2), \nonumber \\
\tilde{\mathbf{q}} &= \frac{p[1+R(\mathbb{M}_1)]\mathbf{q}_1 + (1-p)[1+R(\mathbb{M}_2)]\mathbf{q}_2}{1+r}, \\
\tilde{\mathbb{N}} &= \frac{pR(\mathbb{M}_1) \mathbb{N}_1 + (1-p)R(\mathbb{M}_2)\mathbb{N}_2}{r},
\end{align}
which can straightforwardly be  certified to correspond to a valid probability distribution $\tilde{\mathbf{q}}$ and valid POVM $\mathbb{\tilde{N}}$, we can write
\begin{equation}
p\mathbb{M}_1 + (1-p)\mathbb{M}_2 = (1+\tilde{r})\tilde{\mathbf{q}}\openone - \tilde{r}\tilde{\mathbb{N}}
\end{equation}
as a pseudo-mixture for the convex-combination measurement. The existence of such a pseudo-mixture implies that the robustness can not be larger than $\tilde{r}$, since it provides a feasible solution to \eqref{e:RoM}. It hence follows, as required
\begin{equation}
R(p \mathbb{M}_1 + (1-p)\mathbb{M}_2) \leq p R(\mathbb{M}_1) + (1-p) R(\mathbb{M}_2)
\end{equation}
We note that such a choice for $\tilde{q}$ and $\tilde{\mathbb{N}}$, and hence why only an inequality is obtained (as opposed to an equality). 

Finally, for property (iii), we follow the same logic. In particular, by using the pseudo-mixture for $\mathbb{M}$, we see that
\begin{align}
M'_b &= \sum_a p(b|a) M_a \nonumber \\
&= \sum_a p(b|a) \left( [1+R(\mathbb{M})]q^*(a) \openone - R(\mathbb{M})N^*_a\right) \nonumber \\
&= 	[1+R(\mathbb{M})]\sum_a p(b|a) q^*(a) \openone - R(\mathbb{M}) \sum_a p(b|a)N^*_a
\end{align}
Defining similarly 
\begin{align}
\tilde{q}(b) &= \sum_a p(b|a)q^*(a) \nonumber \\
\tilde{N}_b &= \sum_a p(b|a)N^*_a 
\end{align}
which are seen again to be a valid probability distribution and valid measurement, then 
\begin{equation}
M'_b = [1+R(\mathbb{M})]\tilde{q}(b) \openone - R(\mathbb{M})\tilde{N}_b
\end{equation}
is a valid pseudo-mixture for each element of the measurement, and hence
\begin{equation}
R(\mathbb{M}') \leq R(\mathbb{M})
\end{equation}
as required. 
\subsection{Dual formulation of the RoM}
We start by writing down the Lagrangian associated to the primal SDP \eqref{e:primal}, which is
\begin{align}
\mathcal{L} &= \sum_a \tilde{q}(a) - 1 - \sum_a \tr[\rho_a(\tilde{q}(a)\openone - M_a)] \nonumber \\
&= \sum_a \tr[\rho_a M_a] - 1 + \sum_a \tilde{q}(a)(1 - \tr[\rho_a])
\end{align} 
where we have introduced dual variables (Lagrange multipliers) $\rho_a \geq 0$ which are positive-semidefinite to ensure that the Lagrangian is always smaller than the objective function whenever the constraints of the primal are satisfied. By restricting to dual variables that satisfy $\tr[\rho_a] = 1 $ for all $a$, we see that the Lagrangian becomes independent of the primal variables and equal to $\mathcal{L} = \sum_a \tr[\rho_a M_a] - 1$. Hence, in this case the best upper bound on the objective function is found by maximising over the dual variables and is given by 
\begin{align}
    R(\mathbb{M}) = \max    &\quad \sum_a \tr[\rho_a M_a] - 1\nonumber \\
                \text{s.t.} &\quad \rho_a \geq 0, \quad \tr[\rho_a] = 1 \quad \forall a
\end{align}
which is precisely the dual formulation given in \eqref{e:dual}. Strong duality guarantees that the optimal value of this optimisation problem coincides with the optimal value of the primal problem. Strong duality holds if either the primal or the dual problem are finite and strictly feasible \cite{boyd2004}. It can be seen by inspection that $\rho_a = \openone/d$ for all $a$ constitute a strictly feasible solution to the dual (positive-definite operators), and hence strong duality holds, meaning that \eqref{e:dual} is an equivalent formulation of \eqref{e:primal}. 

\subsection{RoM as advantage in state discrimination}
In this appendix we provide the proof that the Robustness of measurement has operational significance as the best advantage one can obtain in any state discrimination problem over guessing at random, in particular that
\begin{equation}
1 + R(\mathbb{M}) = \max_{\mathcal{E}} \frac{P_\rg^Q(\mathcal{E},\mathbb{M})}{P_\rg^C(\mathcal{E})}
\end{equation}
where
\begin{equation}
P_\rg^Q(\mathcal{E},\mathbb{M}) = \max_{\{P(g|a)\}}\sum_{x,a,g}p(x)\tr[\sigma_x M_a] P(g|a)\delta_{g,x}
\end{equation}
and 
\begin{equation}
p_\rg^C(\mathcal{E}) = p_{\max} = \max_x p(x)
\end{equation}
First we will prove an upper bound on the advantage in any discrimination problem.  From the primal formulation of the RoM \eqref{e:primal} we know that there exist probabilities $\mathbf{q}^* = \{q^*(a)\}$ such that the operator inequality $(1+R(\mathbb{M}))q^*(a)\openone \geq M_a$ is satisfied for all $a$.  This implies that for all discrimination problems 
\begin{align}
P_\rg^Q&(\mathcal{E},\mathbb{M}) \\ 
&\leq (1+R(\mathbb{M})) \max_{\{P(x|a)\}}\sum_{x,a}p(x)q^*(a)\tr[\sigma_x] P(x|a) \nonumber \\
&\leq (1+R(\mathbb{M})) p_{\max}\max_{\{P(x|a)\}}\sum_{x,a}q^*(a)P(x|a) \nonumber \\
&= (1+R(\mathbb{M})) p_{\max}
\end{align}
Since $p_{\max} = p_\rg^C(\mathcal{E})$, we thus see that
\begin{equation}
\frac{P_\rg^Q(\mathcal{E},\mathbb{M})}{P_\rg^C(\mathcal{E})} \leq 1+R(\mathbb{M})
\end{equation}
We note that the right-hand-side is independent of the discrimination problem, and hence provides a bound upon the advantage for all problems. We now wish to show that this bound can be achieved,  i.e. that there is a choice for $\{\sigma_x\}$ and $\mathbf{p}$ that saturate this bound. 

To do so, consider an optimal set of dual variables $\{\rho_a^*\}_a$ for the dual representation of the RoM \eqref{e:dual}. That it, a set of states $\{\rho_a^*\}$ such that $1 + R(\mathbb{M}) = \sum_a \tr[\rho_a^* M_a]$. Let us consider the discrimination problem where the goal is to discriminate between these states, and that they are provided uniformly at random, i.e. $p(x) = 1/o$ for all $x$. That is, let us consider the ensemble $\mathcal{E}^* = \{\rho_x^*, 1/o\}_x$. Let us also assume that the guessing strategy used will be to guess as the state the outcome of the measurement, i.e. $p(g|a) = \delta_{a,g}$ (which might be sub-optimal). Then
\begin{align}
P_\rg^Q(\mathcal{E}^*,\mathbb{M}) &\geq \frac{1}{o}\sum_{x,a,g}\tr[\rho_x^* M_a] \delta_{a,g}\delta_{g,x} \nonumber \\
&= \frac{1}{o}\sum_{x}\tr[\rho_x^* M_x] \nonumber \\
&= p_{\max} (1+R(\mathbb{M}))
\end{align}
where we used the fact that when the states are uniform $p_{\max} = \max_x p(x) = 1/o$. Thus, in this case, we have the advantage 
\begin{equation}
\frac{P_\rg^Q(\mathcal{E}^*,\mathbb{M})}{P_\rg^C(\mathcal{E}^*)} = 1+R(\mathbb{M})
\end{equation}
Thus, the upper bound is achieved precisely by using the states which arise as the solution to the dual SDP \eqref{e:dual}. Altogether, this shows that the RoM has the operational significance of being the advantage, in terms of guessing probability, that the measurement $\mathbb{M}$ provides over guessing purely at random, in an optimally chosen state discrimination problem. 

\subsection{RoM as accessible min-information of a quantum-to-classical channel}
In this appendix we will show that the robustness of measurement can also be seen as equivalent to the accessible min-information of the quantum-classical channel associated to a measurement. In particular, we will show that
\begin{equation}
I_{\min}^\acc(\Lambda_\mathbb{M}(\cdot)) = \log(1 + R(\mathbb{M})),
\end{equation}
where $\Lambda_\mathbb{M}(\rho) = \sum_a \tr[M_a\rho] \ket{a}\bra{a}$ and 
\begin{equation}
I_{\min}^\acc(\Lambda(\cdot)) = \max_{\mathcal{E},\mathbb{D}} I_{\min}(X:G)
\end{equation}
where $I_{\min}(X:G) = H_{\min}(X) - H_{\min}(X|G)$ with $H_{\min}(X) = -\log \max_x p(x) = -\log p_{\max}$ and $H_{\min}(X|G) = -\log \sum_g \max_x p(x,g)$, and $\mathcal{E} = \{\sigma_x,p(x)\}_x$, is the encoding of the classical information $X$ into quantum states $\sigma_x$ which are input into the channel, and  $\mathbb{D} = \{D_g\}_g$ forms a POVM which is measured on the output of the channel and it the decoding of the quantum information back into classical information. Thus, written out explicitly, we have
\begin{align}
I_{\min}^\acc(\Lambda_\mathbb{M}(\cdot)) &= \max_{\{\sigma_x,p(x)\}_x,\{D_g\}_g} -\log p_{\max} + \log \sum_g \max_x p(x,g) \nonumber \\
&= \max_{\{\sigma_x,p(x)\}_x,\{D_g\}_g}  \log \frac{\sum_g \max_x p(x,g)}{p_{\max}} \nonumber \\
&= \max_{\{\sigma_x,p(x)\}_x,\{D_g\}_g}  \log \frac{\sum_g \max_x p(x) \tr[\Lambda_\mathbb{M}(\sigma_x)D_g]}{p_{\max}} \nonumber \\
&= \max_{\{\sigma_x,p(x)\}_x,\{D_g\}_g}  \log \frac{\sum_{g,a} \max_x p(x) \tr[M_a\sigma_x]\bra{a}D_g\ket{a}}{p_{\max}} \nonumber \\
\end{align}
The maximisation over $\{D_g\}_g$ can be evaluated explicitly: Since the output of the channel is already a `classical state' (i.e. a set of orthogonal quantum states), the optimal measurement is to ``read'' this classical register. That is, $D_g = \ket{g}\bra{g}$ is the optimal decoding measurement, hence
\begin{align}
I_{\min}^\acc(\Lambda_\mathbb{M}(\cdot)) &= \max_{\{\sigma_x,p(x)\}_x}  \log \frac{\sum_{g,a} \max_x p(x) \tr[M_a\sigma_x]\delta_{a,g}}{p_{\max}} \nonumber 
\\
&= \max_{\{\sigma_x,p(x)\}_x}  \log \max_{\{P(x|a)\}}\frac{\sum_{a,x} p(x) \tr[M_a\sigma_x]P(x|a)}{p_{\max}} \nonumber 
\end{align}
where in the second line we used the fact that $\max_x f(x) = \max_{\{P(x)\}} \sum_xP(x)f(x)$, with $\{P(x)\}$ an arbitrary probability distribution, to re-express the maximisation over $x$ (for each value of $a$). We finally identify $P_\rg^Q(\mathcal{E},\mathbb{M})=  \max_{\{P(x|a)\}}\sum_{a,x} p(x) \tr[M_a\sigma_x]P(x|a)$ and $P_\rg^C(\mathcal{E}) = p_{\max}$, so that
\begin{align}
I_{\min}^\acc(\Lambda_\mathbb{M}(\cdot)) &= \max_{\{\sigma_x,p(x)\}_x}\log \frac{P_\rg^Q(\mathcal{E},\mathbb{M})}{P_\rg^C(\mathcal{E})} \nonumber \\
&= \log \max_{\{\sigma_x,p(x)\}_x} \frac{P_\rg^Q(\mathcal{E},\mathbb{M})}{P_\rg^C(\mathcal{E})} \nonumber \\
&= \log(1+R(\mathbb{M}))
\end{align}
as required. Thus, the RoM (or a simple function therefore), is equal to the accessible min-information of a quantum-to-classical channel (i.e. a measurement). It thus quantifies, in line with expectation, the maximal amount of correlation (as measure by the min-information) that can be generated by using the channel, between the input information (encoded in the ensemble $\mathcal{E}$), and the outcome of the measurement. 

\subsection{RoA as accessible min-information of an ensemble channel}
In this appendix we show that the Robustness of Asymmetry is also equal to the accessible min-information of a suitably defined ensemble. In particular, we will show that 
\begin{equation}
I_{\min}^\acc(\mathcal{E}_\rho) = \log(1+\mathcal{A}_R(\rho)), 
\end{equation}
where $\mathcal{E}_\rho = \{U_h\rho U_h^\dagger,1/|H|\}$ and
\begin{equation}
I_{\min}^\acc(\mathcal{E}) = \max_{\mathbb{M}} I_{\min}(H:Y)
\end{equation}
As in the previous Appendix, writing everything out explicitly, we have
\begin{align}
I_{\min}^\acc(\mathcal{E}_\rho) &= \max_{\{M_y\}_y} \log |H| + \log \sum_y \max_h p(h,y), \nonumber \\
&= \log |H| +\max_{\{M_y\}_y}  \log \sum_y \max_h \frac{1}{|H|}\tr[U_h\rho U_h^\dagger M_y], \nonumber \\
&= \log |H| +  \log \max_{\{M_y\}_y, \{P(h|y)\}} \sum_{h,y} P(h|y) \frac{1}{|H|}\tr[U_h\rho U_h^\dagger M_y], \nonumber \\
&= \log |H| +  \log \max_{\{M'_h\}_h} \sum_{h} \frac{1}{|H|}\tr[U_h\rho U_h^\dagger M'_h], \nonumber
\end{align}
where to arrive at the third line we used the fact that the maximisation over $h$ inside the sum can be recast as a maximisation over conditional probability distributions, and in the last line we used the fact that an $M_h' = \sum_y P(h|y)M_y$ is an arbitrary measurement when varying over all measurements $\{M_y\}_y$ and all conditional probability distributions $\{P(h|y)\}$. Noticing now that when $\mathcal{E}_\rho = \{U_h \rho U_h^\dagger, 1/|H|\}$ that $\mathbf{q} = \mathbf{1}/|H|$ and $p_\rg^C(\mathbf{1}/|H|) = 1/|H|$, and that $p_\rg^Q(\mathbf{1}/|H|,\rho) = \max_{\{M_h\}_y} \sum_{h} \frac{1}{|H|} \tr[U_h \rho U_h^\dagger M_h]$, we see that
\begin{align}
I_{\min}^\acc(\mathcal{E}_\rho) &= \log\frac{p_\rg^Q(\mathbf{1}/|H|,\rho)}{p_\rg^C(\mathbf{1}/|H|)}, \nonumber \\
&=\log(1+\mathcal{A}_R(\rho)).
\end{align}
That is, the accessible min-information of the ensemble $\mathcal{E}_\rho$ is precisely given by the robustness of asymmetry of the state $\rho$. 
\subsection{Complete set of monotones for post-processing}
In this Appendix we give a complete characterisation of when one measurement can be simulated (via post-processing) by another measurement. In particular, given two measurements $\mathbb{M} = \{M_a\}_a$ and $\mathbb{M}' = \{M'_b\}_b$, we would like to know when it is the case that $M'_b = \sum_a p(b|a)M_a$, for a suitable set of conditional probability distributions $\{p(b|a)\}_{b,a}$. In the main text, we showed that the RoM cannot increase under post-processing, i.e. that $R(\mathbb{M}) \geq R(\mathbb{M}')$ if $\mathbb{M}$ can simulate $\mathbb{M}'$. In the language of resource theories, this shows that the RoM is a monotone. However, there are measurements $\mathbb{M}$ and $\mathbb{M}'$ such that $R(\mathbb{M}) \geq R(\mathbb{M}')$, but such that $\mathbb{M}$ is not able to simulate $\mathbb{M}'$. Thus, knowing only about the RoM of the pair of measurements is not enough to answer the question of whether one can simulate the other or not. 

Our goal here is thus to find a complete set of monotones which, if are all smaller for $\mathbb{M}'$ than $\mathbb{M}$, imply that $\mathbb{M}$ can simulate $\mathbb{M}'$. We will show that such a complete set of monotones exist, and that they are given by the guessing probabilities for any state discrimination game. In particular, we will show that if for all ensembles $\mathcal{E} = \{\sigma_x, q(x)\}_x$ that
\begin{equation}\label{e:monotone}
p_\rg^Q(\mathcal{E},\mathbb{M}) \geq p_\rg^Q(\mathcal{E},\mathbb{M}')
\end{equation}
then $\mathbb{M}$ can simulate $\mathbb{M}'$. That is, $\mathbb{M}$ can simulate $\mathbb{M}'$ if and only if there is never a game where $\mathbb{M}'$ can out-perform $\mathbb{M}$ in the task of guessing which state from the ensemble was prepared on average. 

To prove this claim, we start with the easy direction, namely that if $\mathbb{M}'$ can be simulated by $\mathbb{M}$ then it can never outperform it in any state discrimination game. This follows immediately, as can be seen. In particular, if $\mathbb{M} = \{M_a\}_a$ can simulate $\mathbb{M}' = \{M'_b\}_b$, then there exists a collection of probability distributions $\{p(b|a)\}_{ab}$ such that 
\begin{equation}
M_b' = \sum_a p(b|a)M_a.
\end{equation}
Then, for any given discrimination game specified by an ensemble $\mathcal{E} = \{\sigma_x, q(x)\}_x$, the success probability using $\mathbb{M}'$ is 
\begin{align}
p_\rg^Q(\mathcal{E},\mathbb{M}') &= \max_{\{P'(g|b)\}} \sum_{b,x,g} q(x) \tr[\sigma_x M'_b] P'(g|b) \delta_{g,x}, \nonumber \\
&= \max_{\{P'(g|b)\}} \sum_{b,x,g,a} q(x) p(b|a)\tr[\sigma_x M_a] P'(g|b) \delta_{g,x}, \nonumber \\
&\leq \max_{\{P(g|a)\}} \sum_{x,g,a} q(x)\tr[\sigma_x M_a] P(g|a) \delta_{g,x}, \nonumber \\
&= p_\rg^Q(\mathcal{E},\mathbb{M})
\end{align}
where in the third line we used the fact that $\sum_b p(b|a)P'(g|b) = P(g|a)$ forms a conditional probability distribution, and that, depending on $p(b|a)$, this might not be the most general set of conditional probability distributions, which leads to the inequality. This proves the ``if'' direction. Let us now prove the ``only if'' direction.

Let us therefore assume that $\mathbb{M}'$ and $\mathbb{M}$ satisfy \eqref{e:monotone} for all $\mathcal{E}$. Written in full, that is
\begin{multline}
\max_{\{P(x|a)\}} \sum_{a,x} q(x) \tr[\sigma_x M_a] P(x|a) \\ - \max_{\{P'(x|b)\}} \sum_{x,b} q(x) \tr[\sigma_x M'_b] p'(x|b) \geq 0 
\end{multline}
Let us assume (a potentially sub-optimal) choice $P'(x|b) = \delta_{x,b}$ (i.e. we guess the state was $\sigma_b$ when we see the result of the measurement is $b$) then we also have
\begin{align}
&\max_{\{P(x|a)\}} \sum_{a,x} q(x) \tr[\sigma_x M_a] P(x|a) - \sum_{x,b} q(x) \tr[\sigma_x M'_b] \delta_{x,b}   \nonumber \\
&= \max_{\{P(x|a)\}} \sum_{x} q(x) \tr\left[\sigma_x \left(\sum_a P(x|a)M_a - M'_x\right)\right] \geq 0
\end{align}
Since this holds for all $\mathcal{E}$, it also holds if we minimise over all $\mathcal{E}$, and therefore
\begin{equation}
\min_{\mathcal{E}}\max_{\{P(x|a)\}} \sum_{x} q(x) \tr\left[\sigma_x \left(\sum_a P(a|x)M_a - M'_x\right)\right] \geq 0
\end{equation}
Now, since the function being optimised is linear in $P(x|a)$ and linear in $\tilde{\sigma_x} = q(x)\sigma_x$, this implies we can invoke the minimax theorem  and reverse the order of the minimisation and maximisation, leading to
\begin{equation}\label{e:starting condition}
\max_{\{p(x|a)\}} \min_{\mathcal{E}} \sum_{x} q(x) \tr\left[\sigma_x \left(\sum_a p(x|a)M_a - M'_x\right)\right] \geq 0
\end{equation}
Consider first that a collection of conditional probabilities $\{P(x|a)\}$ exist such that $\sum_a P(x|a)M_a - M'_x = 0$ for all $x$, i.e. such that $\mathbb{M}$ can simulate $\mathbb{M}'$. In this case, the objective function vanishes for all $\mathcal{E}$ and we satisfy the inequality. Let us now assume that no such $\{P(x|a)\}$ exists, and show that this leads to a contradiction. 

If no such set of probabilities exists, then the Hermitian operators $\Delta_x := \sum_a P(x|a)M_a - M'_x$ cannot be identically equal to the zero-operator for all values of $x$. However, by nature of being conditional probability distributions, we have nevertheless $\sum_x P(x|a) = 1$ for all $x$, and hence
\begin{equation}
\sum_x \Delta_x = \sum_x \sum_a P(x|a)M_a - M'_x = 0 
\end{equation}
This in turn implies that it is impossible that all $\Delta_x \leq 0$, i.e. are negative semidefinite operators. Indeed, the sum of a set of negative semidefinite operators can only equal the zero-operator if each operator is itself the zero-operator, which we take not to be the case by assumption. Therefore, there exists at least one choice $x^*$ such that $\Delta_x^*$ has at least one negative eigenvalue $\lambda_\text{neg}^{x^*}$. Let us denote the associated eigenvector by $\ket{\lambda_\text{neg}^{x^*}}$. Finally, let us then choose as an ensemble $\mathcal{E}$ one such that $q(x^*) = 1$ and $\sigma_{x^*} = \ket{\lambda_\text{neg}^{x^*}}\bra{\lambda_\text{neg}^{x^*}}$. We then find that
\begin{equation}
\sum_{x} q(x) \tr\left[\sigma_x \left(\sum_a p(x|a)M_a - M'_x\right)\right] = \lambda_\text{neg}^{x^*} < 0
\end{equation}
which is in contradiction with \eqref{e:starting condition}. Thus, we conclude that our assumption cannot hold, and it must be the case that $\Delta_x = 0$ for all $x$, which is equivalent to $\mathbb{M}$ being able to simulate $\mathbb{M}'$. This thus proves that simulability is equivalent to never being more useful in any discrimination game. 

We will finish by making two remarks about this result. First, we expressed it in terms of the guessing probability, but we can also re-phrase it in terms of conditional min-entropies. In particular, implicit in our previous analysis was the fact that
\begin{equation}
H_{\min}(X|A)_{\mathcal{E},\mathbb{M}} = -\log p_\rg^Q(\mathcal{E},\mathbb{M})
\end{equation} 
where
\begin{equation}
p(a,x) = q(x) \tr[\sigma_x M_a]
\end{equation}
is the joint probability distribution induced by the ensemble $\mathcal{E}$ and measurement $\mathbb{M}$. With this defined, we see that a measurement $\mathbb{M}$ can simulate a measurement $\mathbb{M}'$ if and only if
\begin{equation}
H_{\min}(X|A)_{\mathcal{E},\mathbb{M}} \leq H_{\min}(X|A)_{\mathcal{E},\mathbb{M'}} \quad \forall \mathcal{E}.
\end{equation}
This form for the complete set of monotones makes the connections to previous results more direct \cite{buscemi2016,gour2017b}. 

Second, it is worth noting how this complete characterisation implies the previously obtained result that $R(\mathbb{M}) \geq R(\mathbb{M}')$ whenever $\mathbb{M}$ can simulate $\mathbb{M}'$, which was proved independently. This follows from the following sequence,
\begin{align}
1 + R(\mathbb{M}) &= \frac{p_\rg^Q(\mathcal{E}^*,\mathbb{M})}{1/o_\mathbb{M}} \nonumber \\
&\geq \frac{p_\rg^Q(\mathcal{E'}^*,\mathbb{M})}{1/o_{\mathbb{M}'}} \nonumber \\
&\geq \frac{p_\rg^Q(\mathcal{E'}^*,\mathbb{M}')}{1/o_{\mathbb{M}'}} = 1 + R(\mathbb{M}')
\end{align}
where the equality in the first line follows from the operational significance of the RoM as the optimal advantage for any discrimination game specified by an ensemble $\mathcal{E}$, \eqref{e:op sig RoM}, and where we have defined the optimal ensemble $\mathcal{E}^*$ for the measurement $\mathbb{M}$ with $o_\mathbb{M}$ outcomes. The second inequality follows by the fact that the advantage for the optimal ensemble is not smaller than the advantage for the ensemble $\mathcal{E'}^*$ (with $o_{\mathbb{M}'}$ outcomes) which is the optimal ensemble for the measurement $\mathbb{M}'$. The third inequality uses \eqref{e: monotone}, i.e. that simulability implies that the guessing probability cannot be larger for any ensemble, and then the final equality uses again \eqref{e:op sig RoM}.

\end{appendix}
\end{document}